\pdfoutput=1

\documentclass[prl,reprint,twocolumn,showpacs,preprintnumbers,amsmath,amssymb,superscriptaddress,nofootinbib]{revtex4-1}
\usepackage{graphicx}

\usepackage{epsfig}
\usepackage{color}
\usepackage{amssymb}
\usepackage{amsmath}
\usepackage{mathrsfs}
\usepackage{hyperref}
\usepackage{multirow}

\def\be{\begin{equation}}
\def\ee{\end{equation}}
\def\ba{\begin{eqnarray}}
\def\ea{\end{eqnarray}}
\def\nl{\nonumber\\}
\def\Li{\textrm{Li}}

\def\eps{\epsilon}

\def\EE{\mathcal{E}}
\def\Et{\tilde{E}}

\def\Gcusp{\Gamma_{\rm cusp}}

\begin{document}

\null\vskip-10pt \hfill
\begin{minipage}[t]{42mm}
SLAC--PUB--16811\\
\end{minipage}
\vspace{0mm}

\title{Bootstrapping a Five-Loop Amplitude Using Steinmann Relations}
\author{Simon Caron-Huot}\affiliation{Niels Bohr International Academy \& Discovery Center, Niels Bohr Institute, University of Copenhagen, Denmark}
\affiliation{Department of Physics, McGill University, 3600 Rue University, Montr\'eal, Qu\'ebec, Canada H3A 2T8} 
\author{Lance J. Dixon}\affiliation{SLAC National Accelerator Laboratory, Stanford University, Stanford, CA 94309, USA}
\affiliation{Kavli Institute for Theoretical Physics, UC Santa Barbara, Santa Barbara, CA 93106, USA}
\author{Andrew McLeod}\affiliation{SLAC National Accelerator Laboratory, Stanford University, Stanford, CA 94309, USA}
\author{Matt von Hippel}\affiliation{Perimeter Institute for Theoretical Physics, Waterloo, Ontario, Canada}
\date{\today}

\begin{abstract}
The analytic structure of scattering amplitudes is restricted by Steinmann relations, which enforce the vanishing of certain discontinuities of discontinuities.
We show that these relations dramatically simplify the function space
for the hexagon function bootstrap in planar maximally supersymmetric Yang-Mills theory. Armed with this simplification, along with the constraints of dual conformal symmetry and Regge exponentiation,
we obtain the complete five-loop six-particle amplitude.
\end{abstract}

\maketitle



\section{Introduction}\label{sec:intro}\vspace{-10pt}

To ``bootstrap'' generally refers to solving a problem via an ansatz constrained by symmetries and physical principles.
This is naturally most successful in very special theories such
as low-dimensional integrable models, but it has also proved powerful for conformal field theories in arbitrary dimensions.
The hexagon function bootstrap~\cite{Dixon2011pw,Dixon2015iva}
is a perturbative version aimed at solving a scattering problem
in a four-dimensional quantum field theory:
the planar limit of $\mathcal{N}=4$ super Yang-Mills (SYM).
While scattering amplitudes in this theory are interesting in their own
right, the methods developed to solve them have often had broader applicability,
for example to computing amplitudes in QCD for scattering
at the Large Hadron Collider.

The hexagon function bootstrap exploits the idea that, order by order in 
perturbation theory, the first nontrivial 
amplitude in planar $\mathcal{N}=4$ SYM, the six-point amplitude,
``lives'' within a relatively small space of functions, which
can be parametrized by a finite set of coefficients.
This rigidity means that information from physical limits, such as when
two gluons become collinear, or in a high-energy (Regge) limit,
often suffices to fix the result.  In turn this generates new predictions,
a fact which has led to much fruitful interplay with the pentagon
operator-product-expansion
program~\cite{Basso2013vsa,Basso2013aha,BelitskyI,POPE}.

The aim of this Letter is to point out that the relevant space of hexagon
functions is far smaller than previously thought.
This is due to constraints stemming from the classic work of Steinmann~\cite{Steinmann},
which restrict the analytic structure of scattering amplitudes in any quantum field theory.
We show that, when combined with Regge exponentiation and the so-called final-entry condition~\cite{CaronHuot2011kk},
this restriction makes it possible to bootstrap the six-gluon amplitude to at least 5 loops \emph{without any external input}.  Analogous constraints can be exploited for $n$-particle scattering with $n>6$.

\section{Hexagon Steinmann functions}

We consider the scattering amplitude for six gluons (or other partons) in
the planar limit of $\mathcal{N}=4$ SYM.
A priori, such an amplitude can depend, in four spacetime dimensions,
on 8 Mandelstam invariants.  Dual conformal symmetry of this model 
restricts the nontrivial dependence to be on 3 
cross-ratios~\cite{DualConformal,RemainderFunction}
\be
 u=\frac{s_{12}s_{45}}{s_{123}s_{345}},\quad
 v=\frac{s_{23}s_{56}}{s_{234}s_{123}},\quad
 w=\frac{s_{34}s_{61}}{s_{345}s_{234}}, \label{cross-ratios}
\ee
where $s_{i\ldots k}\equiv (p_i{+}\cdots{+}p_k)^2$ are Mandelstam invariants.
The same symmetry forces the four- and five-particle amplitudes to be
essentially trivial, which is why we concentrate on six particles.
It has been conjectured that the amplitude,
which is a transcendental function of these three variables,
lives in a restricted space of ``hexagon'' functions~\cite{Dixon2011pw}.
These are iterated integrals with singularities generated by logarithms
of the nine letters~\cite{Goncharov2010jf}
\be
{\cal S}\ =\ \{ u,v,w,1{-}u,1{-}v,1{-}w,y_u,y_v,y_w\}, \label{alphabet}
\ee
where
\be
 y_u = \frac{1{+}u{-}v{-}w{-}\sqrt{\Delta}}{1{+}u{-}v{-}w{+}\sqrt{\Delta}},
\quad
\Delta=(1{-}u{-}v{-}w)^2-4uvw,\nonumber
\ee
and cyclic rotations act as
\be
C:\quad u\to v\to w\to u,\quad y_u\to 1/y_v\to y_w\to 1/y_u,
\label{Cdef}
\ee
while parity acts as $u_i\to u_i$, $y_i\to 1/y_i$.
These letters arise naturally as projectively invariant combinations of
momentum twistors~\cite{Hodges2009hk}, variables that make manifest the
dual conformal symmetry.  Multiple zeta values $\zeta_{q_1,q_2,\ldots}$
with positive indices $q_i$ also appear.

Branch cuts for massless scattering amplitudes start only at vanishing
values of the Mandelstam invariants, $s_{i\ldots k}=0$.  Consequently,
there is a canonical Riemann sheet on which the amplitude is
analytic in the positive octant $u,v,w>0$.
This constraint is included in the definition of hexagon functions.
It implies a ``first-entry'' condition~\cite{Gaiotto2011dt}:
discontinuities associated with the letters $(1{-}u)=0$ or $y_u=0$ are
not visible in the canonical Riemann sheet; however, they
can be exposed after analytic continuation.
The physical interpretation of the restriction~(\ref{alphabet}) is that,
even after analytic continuation along an arbitrary complex path,
the only possible branch points remain those characterized by ${\cal S}$.

\begin{figure}
\includegraphics[width=2.5in]{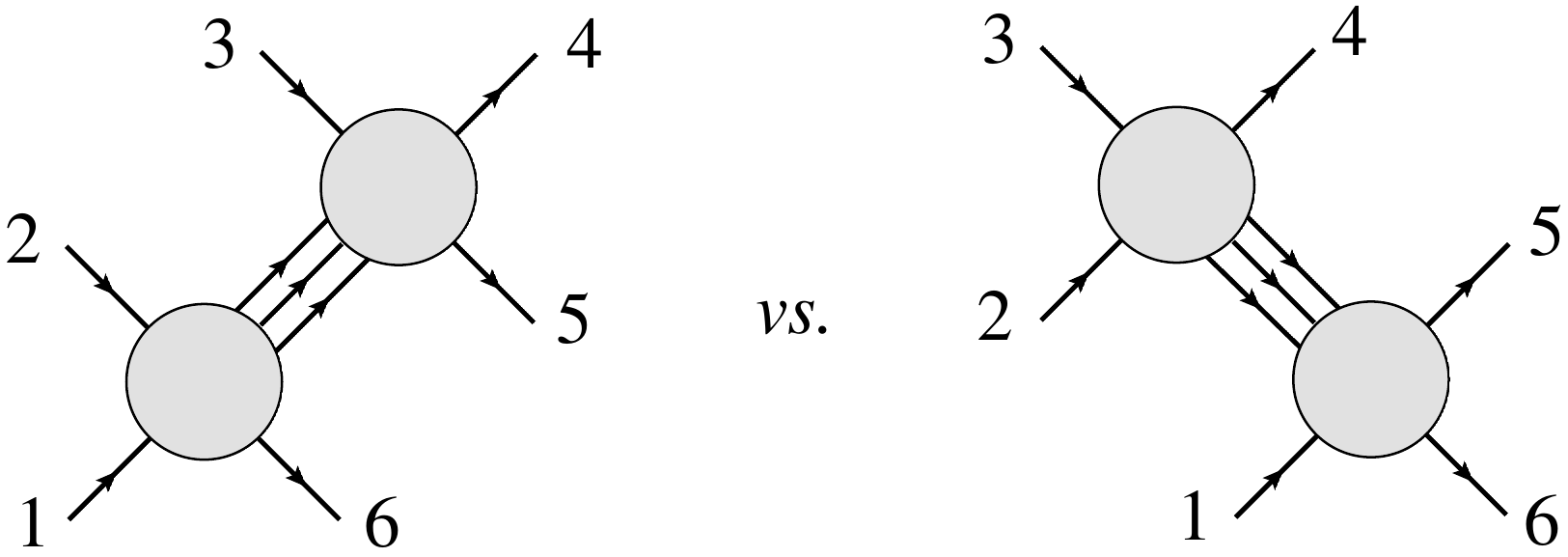}
\caption{Illustration of the channels $s_{345}$ and $s_{234}$ for $3\to3$
kinematics. The discontinuity in one channel should not know about
the discontinuity in the other channel.}
\label{fig:stein}
\end{figure}

The focus of this Letter is the Steinmann relations, which state that
an amplitude $A$ can have no double discontinuities in overlapping
channels~\cite{Steinmann}.  Using the correspondence between discontinuities
and cut diagrams via the Cutkosky rules~\cite{Cutkosky1960sp},
overlapping channels correspond to cut lines that intersect.
Thus for example the channels $s_{345}$ and $s_{234}$ overlap,
which leads, schematically, to:
\be
\mbox{Steinmann relation:}\quad {\rm Disc}_{s_{345}} \left({\rm Disc}_{s_{234}} A\right) =0, \label{Steinman_v0}
\ee
illustrated in figure \ref{fig:stein}.

We focus on three-particle invariants $s_{ijk}$ because these can change sign along fairly generic codimension-1 surfaces
in the space of external momenta.  The relation can therefore be probed with real external momenta.
(In contrast, massless thresholds in two-particle invariants $s_{ij}$ occur at phase space boundaries where other invariants may change sign; it is unclear to the authors how to extract putative constraints from these thresholds beyond the Regge limit~\cite{Bartels2008ce}.)
For functions of the cross-ratios $u,v,w$, the discontinuity with
respect to $s_{234}$ can be computed by rotating $v,w$ by a common phase,
as follows from eq.~(\ref{cross-ratios}).
The general Steinmann relation~(\ref{Steinman_v0})
thus implies --- for the special case of dual-conformally
invariant functions --- that the following combination
is analytic in a neighborhood of $r=\infty$:
\ba
0= {\rm Disc}_{r=\infty} \big[
       A(r u, v e^{i\pi}, r e^{i\pi})
     - A(r u, v e^{-i\pi}, r e^{-i\pi}) \big],\nl
\label{Steinmanndc}
\ea
where $u,v>0$ (and $r>0$ before taking the discontinuity).
The reason why $r=\infty$ appears is that the three-particle invariants appear in the denominators of eq.~(\ref{cross-ratios}).

Focusing on the region where all three cross-ratios are large and
combining this condition with its permutations,
we obtain an equivalent but more practical statement:
the amplitude must be expressible as a sum of terms with singularities in only one three-particle channel:
\be
 A = \sum_k \left[a_k^u\log^k \left(\frac{u}{vw}\right)+a_k^v\log^k \left(\frac{v}{wu}\right)+a_k^w\log^k \left(\frac{w}{uv}\right)\right], \label{Steinman_v2}
\ee
with the $a_k^{u,v,w}$ analytic around $u=v=w=\infty$.


\section{The Steinmann basis to weight 4}

A complete basis of 88 hexagon functions at transcendental weight 4 was
originally constructed in ref.~\cite{Dixon2011nj}.
The Steinmann relations imply that only a subspace is physically relevant,
a subspace sufficiently small that it can be described in this Letter.
We begin with weight 1, where the first entry condition allows only
elementary logarithms: $\log u$, $\log v$, $\log w$.
To build the higher weight basis, we use the fact that all derivatives
of a Steinmann function also obey the Steinmann relations.

The derivative of a weight-$k$ hexagon function $F$ has the
form~\cite{Dixon2013eka}
\be
dF = \sum_{i=1}^9 F^i \ d\ln{\cal S}_i \,,
\label{dF}
\ee
where $F^i$ are weight-$(k{-}1)$ hexagon functions and
${\cal S}_i\in{\cal S}$ in eq.~(\ref{alphabet}).
We thus make an ansatz~(\ref{dF}) for the derivatives of $F$
where the $F^i$ are Steinmann functions.
For the ansatz to represent a function, the partial derivatives must commute
(``integrability condition'').
Once this condition is solved, the analyticity and Steinmann properties
simplify dramatically. It suffices to impose the
following constraints, which serve only to fix
a few coefficients of zeta-values of weight $(k{-}1)$ and $(k{-}2)$:
\begin{itemize}
\item $F^{1{-}u}$, $F^{y_v}$ and $F^{y_w}$
must vanish at $(u,v,w)=(1,0,0)$~\cite{Dixon2013eka,Dixon2015iva}.
\item The $s_{234}$-discontinuity of $F^u + F^{1-u} + F^w + F^{1-w}$
must vanish at $(u,v,w)=(+\infty,0,-\infty)$.
\end{itemize}
Cyclic rotations of these conditions are implied.
The first condition enforces the absence of unwanted
discontinuities~\cite{Gaiotto2011dt} at function level;
the second condition does the same for the Steinmann
condition~(\ref{Steinmanndc}).

Following this procedure, at weight 2 we find 7 elements:
the constant $\zeta_2$ and two cyclic orbits containing
\ba
K_{1,1}^u \equiv\Li_2(1{-}1/u),\quad 
L_2^u\equiv \tfrac12\!\left[\log^2(u)+\log^2(v/w)\right].\nl
\label{weight_2}
\ea
The naming convention will be explained shortly.
Already, the Steinmann relations' impact is noticeable:
without it there would be three additional functions,
$\log^2u$, $\log^2v$ and $\log^2w$,
which do not satisfy eq.~(\ref{Steinman_v2}).

At weight 3, the basis contains 17 elements, the 5 cyclic 3-orbits of
\ba
 K^u_{3}&\equiv&\tfrac{1}{3!}\log^3(1/u) + \tfrac12\log(1/u)\log^2(v/w),\nl
 K_{2,1}^u &\equiv& \Li_2(1/u) \log(1/u)-2\Li_3(1/u)+2\zeta_3, \nl
 K_{1,2}^u &\equiv& K_{1,1}^u\log(v/w),\qquad
 K_{1,1,1}^u\equiv -\Li_3(1{-}1/u),  \nl
 \zeta_2K_{1}^u&\equiv& \zeta_2\log(1/u),
\label{weight_3}
\ea
the constant $\zeta_3$, and a single parity-odd element:
the six-dimensional scalar hexagon integral
$\tilde\Phi_6$~\cite{Dixon2011ng,Dixon2013eka}.

At this stage we see that the functions in 
eqs.~(\ref{weight_2})-(\ref{weight_3}) depend nontrivially on only $u$,
apart from simple powers of $\log(v/w)$.
We can construct $3\times 2^{k{-}1}$ similar elements at weight $k$, as follows.
We start from ``seeds'' which trivially satisfy eq.~(\ref{Steinman_v2}):
\be\begin{aligned}
 K^u_{k}(u,\tfrac{v}{w}) &\equiv \frac1{2\cdot k!}\left[ \log^k\left(\frac{v}{uw}\right)-\log^k\left(\frac{uv}{w}\right)\right],\\
 L^u_{k}(u,\tfrac{v}{w}) &\equiv \frac1{2\cdot k!}\left[ \log^k\left(\frac{v}{uw}\right)+\log^k\left(\frac{uv}{w}\right)\right].
\end{aligned}\ee
We then construct nontrivial functions as a simple generalization of harmonic polylogarithms (HPLs)~\cite{Remiddi1999ew} with argument $x=1/u$,
by integrating the seeds from the base point $u=\infty$.  Using this base point automatically maintains the Steinmann relations.
The constraint of analyticity for $u>0$ is enforced by recursively removing values at $u=1$:
\be
  K^u_{i,\ldots}(u,\tfrac{v}{w}) \equiv \sum_j c_j L^u_j+\int_0^{1/u}\!\! \frac{dx}{1-x}
  \frac{\log^{i{-}1}\!(\tfrac{1}{ux})
  }{(i-1)!}
  K^u_{\ldots}(\tfrac{1}{x},\tfrac{v}{w}),
\ee
where the zeta-valued coefficients $c_j$ are chosen uniquely to make the total vanish at $u=1$.
Without the $c_j$, the recursive definition would be identical to that of HPLs with argument $x=1/u$,
which makes it straightforward to express the $K^u$ as combinations of HPLs.
At weights 2 and 3, this definition agrees with the examples given.

Defining $K^v$, $K^w$, $L^v$ and  $L^w$ as cyclic images of $K^u$, $L^u$,
the $K$ functions with positive indices do generate
$3\times 2^{k{-}1}$ linearly independent elements.
There is one exception: the three $K^{u,v,w}_{k}$ for even weight $k$ are
linearly dependent, so for even $k$ we use $L_k^{u,v,w}$ instead.  

At weight 4, the Steinmann basis contains the 8 3-orbits
generated by:
\def\hquad{\hspace{1.7mm}}
\be
L_4^u, \hquad K_{1,3}^u,\hquad K_{2,2}^u,\hquad K_{3,1}^u,\hquad 
K_{1,1,2}^u,\hquad K_{1,2,1}^u,\hquad K_{2,1,1}^u,\hquad K_{1,1,1,1}^u.
\nonumber
\ee
The iterative construction also generates 5 ``non-$K$'' functions: 
3 parity-even functions
--- the integral $\Omega^{(2)}$~\cite{Dixon2011nj,Dixon2013eka}
and its cyclic permutations --- plus 2 parity-odd functions.
Ten more functions come from multiplying $\zeta_2$, $\zeta_3$ and $\zeta_4$
by the lower-weight Steinmann functions listed earlier.
In summary, at weight 4 there are 39 physically relevant Steinmann functions,
to be contrasted with 88 in the original hexagon function space.

This gap increases rapidly with higher weights, as evidenced by the
first two lines of table \ref{tab:constraints},
which was generated by implementing the construction iteratively.
The paucity of Steinmann functions is because the space is
not a ring: the product of two Steinmann functions is 
generically \emph{not} an allowed function.


\renewcommand{\arraystretch}{1.25}
\begin{table}[!t]
\centering
\def\sp{@{{\,}}}
\begin{tabular}[t]{l\sp c\sp c\sp c\sp c\sp c\sp c\sp c\sp c}
\hline
Constraint                      & $L=1$ & $L=2$ & $L=3$ & $L=4$ & $L=5$
\\\hline
0. Functions    & (10,10) & (82,88) & (639,761) & (5153,6916)\! & (???,???)\\\hline
1. Steinmann       & (7,7) & (37,39) & (174,190) &  (758,839) & \!(3105,3434)\\\hline
2. Symmetry      & (3,5) & (11,24) &  (44,106) &  (174,451) & (???,???)\\\hline
3. Final-entry\!\! & (2,2) & (5,5) &   (19,12) &   (72,32) & (272,83)\\\hline
4. Collinear               & (0,0) & (0,0) &  (1,1) &    (3,5) &   (9,15)\\\hline
5. Regge                            & (0,0) & (0,0) &    (0,0) &    (0,0) &   (0,0)\\\hline
\end{tabular}
\caption{Free parameters remaining after applying each constraint,
for the 6-point (MHV,NMHV) amplitude at $L$ loops.}
\label{tab:constraints}
\end{table}

\section{Application to two loops}

Before using the Steinmann basis to help bootstrap the hexagon amplitude,
we comment on the subtraction of its infrared divergences.
A particularly convenient
scheme for removing infrared divergences in the SYM model is to
divide by the so-called BDS ansatz~\cite{BDS}. This soaks up the dual
conformal anomaly, leaving a remainder which depends only on the
cross-ratios $u,v,w$, and furthermore vanishes in soft and collinear
limits~\cite{RemainderFunction}.\footnote{The reader may object
that higher-order poles in $\eps=(4-D)/2$ in the BDS ansatz mean that
the full amplitude is not determined through ${\cal O}(\eps^0)$
by the remainder function alone.  However, it has been proved~\cite{Weinzierl}
at next-to-next-to-leading order that the higher-order terms in $\eps$
in one-loop amplitudes are not needed, if one knows the
two-loop remainder function to ${\cal O}(\eps^0)$.  Based on the universal
nature of infrared divergences and their cancellation,
we expect the same result to hold to higher perturbative orders.}

However, in order to preserve the Steinmann relation (\ref{Steinman_v0}),
it is critical to divide only by quantities which are free
of three-particle discontinuities. This singles out the so-called BDS-like
ansatz~\cite{Alday2009dv,Dixon2015iva} $\mathcal{R}_6'$:
\be
 \mathcal{R}_6'\equiv \mathcal{M}^{\rm bare}_6 / \mathcal{M}_6^{\rm BDS-like} \,.
\ee
In fact, the amplitude is a function of the helicity of all 6 particles,
in a way which can be neatly
encoded in so-called $R$-invariants~\cite{Drummond2008vq,Hodges2009hk}.
In this Letter we thus deal with bosonic functions $\EE$, $E$ and $\tilde E$ which encode all the information
and correspond to suitable components of the MHV and NMHV BDS-like remainders. 
Schematically, $\mathcal{R}_6'\simeq \EE\oplus E \oplus \Et$. 
The relations to the more conventional BDS MHV remainder ($\mathcal{R}_6$)
and NMHV ratio function ($V,\tilde{V}$),
defined for example in ref.~\cite{Dixon2015iva}
(to which we refer for further details), are:
\be
 e^{\mathcal{R}_6}\equiv \EE e^{-\frac14\Gcusp\mathcal \EE^{(1)}},\quad
 V\equiv E/\EE,\quad \tilde{V}\equiv \Et/\EE, \label{relations_to_standard}
\ee
where $\tfrac14\Gcusp=g^2-2\zeta_2g^4+\ldots$ is the cusp anomalous dimension,
known exactly as a function of the coupling
$g^2\equiv \frac{g^2_{\rm YM}N_c}{16\pi^2}$~\cite{Beisert2006ez}.
We stress that while $\EE$, $E$ and $\Et$ obey the Steinmann
relations, $\mathcal{R}_6$, $V$ and $\tilde{V}$ do not:
the space of Steinmann functions is not a ring.

Let us describe a concrete example, the bootstrap of $\EE$ at two loops.
We begin by applying the following:
\begin{itemize}
\item[1.] $\EE$ is a hexagon Steinmann function
\item[2.] $\EE$ is parity-even and dihedrally symmetric
\item[3.] The collinear limit to leading power is universal:
\be
 \lim_{v\to 0} \EE = e^{-\frac14\Gcusp(L_2^v + 2\zeta_2)}
 + {\cal O}(\sqrt{v} \, \ln^{L-1} v).\nonumber
\ee
\end{itemize}
In the weight 4 Steinmann space, no linear combination
vanishes in all three collinear limits.  Therefore the two-loop MHV 
amplitude is fully determined by just the above three conditions! 
Loop expanding using
$\EE = \EE^{(0)} + g^2 \EE^{(1)} + g^4 \EE^{(2)} + \ldots$,
the result at tree level is $\EE^{(0)}=1$, at one loop
\be
 \EE^{(1)} = K_{1,1}^u+K_{1,1}^v+K_{1,1}^w \,,
\ee
and at two loops
\be
 \EE^{(2)} {=} (1{+}C{+}C^2)\big[
 \Omega^{(2)}{-}K_{1,2,1}^u{-}4K_{1,1,1,1}^u{-}\zeta_2 K_{1,1}^u\big]{+}8\zeta_4 \,,
\ee
where the cyclic rotation $C$ is defined in eq.~(\ref{Cdef}).
This result agrees completely with refs.~\cite{Goncharov2010jf,Dixon2011nj}.

For MHV at higher loops, and for NMHV, we imposed
an additional ``final-entry'' condition, obtained
by considering the action of the $\bar Q$ generator of dual 
superconformal transformations~\cite{CaronHuot2011kk}.
The MHV final-entry condition is simply $\EE^{1-u} = - \EE^u$, 
plus the cyclic relations.
Similarly, the differential of the NMHV BDS-like remainder is spanned
by the 18 elements listed in eq.~(3.10) of ref.~\cite{Dixon2015iva}.
These conditions almost completely determine the higher-loop amplitudes;
we need information from only one more limit.


\section{Regge exponentiation and bootstrap}

In the multi-Regge limit of $2\rightarrow 4$ gluon scattering,
the four outgoing gluons are strongly ordered in rapidity.  The cross-ratios
have the limits $u\rightarrow 1$, $v,w\rightarrow 0$,
but on an analytically continued Riemann sheet which ensures nontrivial
Lorentzian kinematics.
This limit has been thoroughly analyzed for both MHV and NMHV
amplitudes~\cite{Bartels2008ce,Bartels2008sc,Fadin2011we,Lipatov2012gk,BCHS,Dixon2014iba}.
Amplitudes exponentiate in terms of Fourier-Mellin variables $\nu,m$
which are conjugate to the transverse plane coordinates, schematically:
\be
\EE(\nu,m,vw)\xrightarrow{{\rm Regge}}\Phi(\nu,m)\times \left(-1/\sqrt{vw}\right)^{\omega(\nu,m)}
\ee
where the Regge trajectory $\omega$ vanishes at tree level and $\Phi$ is an ``impact factor''.
Exponentiation implies that terms with $\log^2(vw)$ or higher powers
of the large logarithm are predicted by the multi-Regge limit at lower loops.

Remarkably, through five loops such terms suffice to fix all remaining parameters and
uniquely determine $\EE$, $E$, and $\Et$! Terms with $\log(vw)$ or lower were not needed, but rather
led to predictions for the next loop order, enabling a pure bootstrap with no external information.
The constraints are summarized in table \ref{tab:constraints}.

\begin{figure}
\includegraphics[width=3.5in]{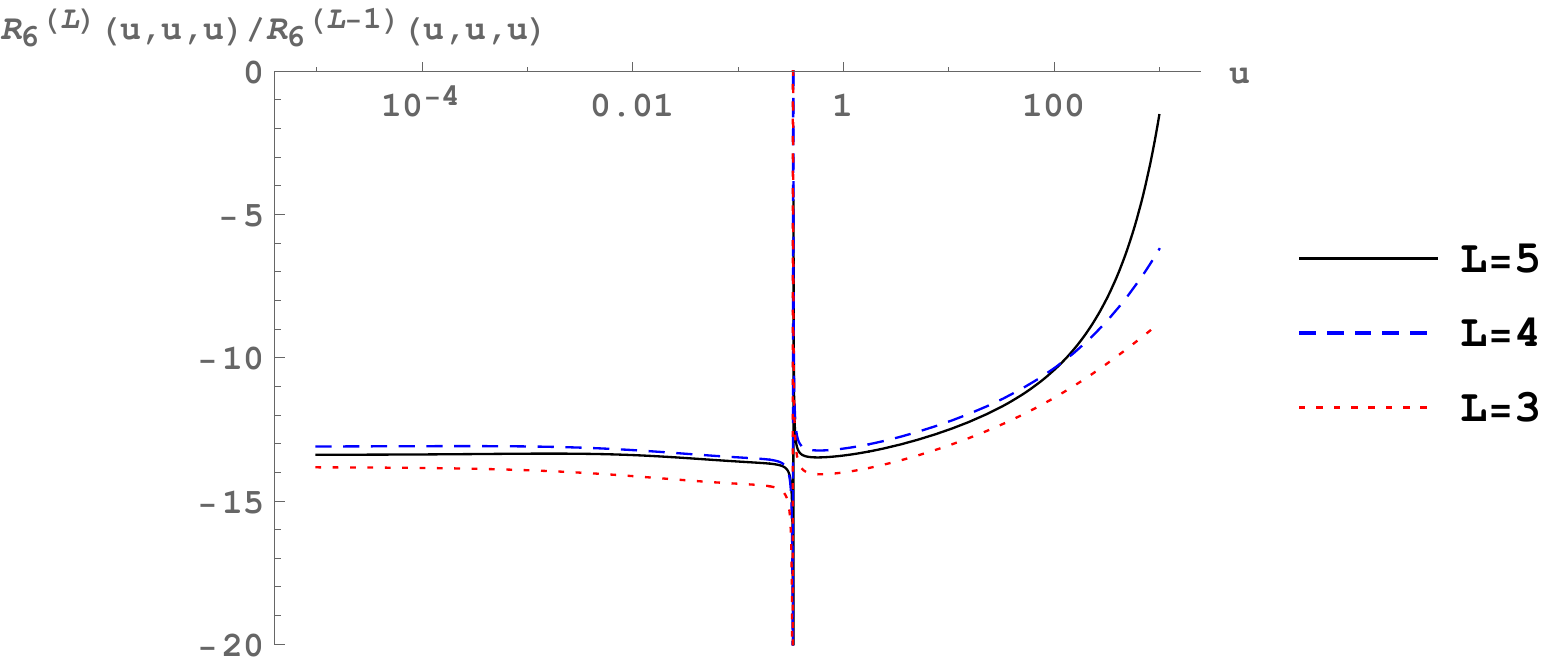}
\caption{The remainder function $\mathcal{R}_6$,
evaluated at ratios of successive loop orders $L$ on the line $u=v=w$.
The spike is an artifact due to $\mathcal{R}_6^{(L)}(u,u,u)$
crossing zero very close to $u=1/3$ at each loop order.}
\label{fig:ratios}
\end{figure}

With $\EE$, $E$, and $\Et$ fixed through five loops we can evaluate them numerically on a variety of lines in cross-ratio space. Figure \ref{fig:ratios} shows the remainder function on the line $(u,u,u)$. We have also used ``hedgehog'' variables~\cite{Parker2015cia} to generate multiple polylog representations of these functions in one bulk region~\cite{SteinmannWebsite}.

Past implementations of the hexagon function bootstrap employed
a variety of other constraints, which the Steinmann relations
render unnecessary, or relegate to cross checks.
For NMHV, the representation in terms of $R$-invariants
has poles at kinematically spurious points that must cancel
between different permutations of $E$ and $\tilde{E}$~\cite{Dixon2011nj}.
Now, after imposing the collinear constraint in 
table \ref{tab:constraints}, the spurious poles cancel automatically.
Similarly, for MHV and NMHV the $\bar{Q}$ equation predicts not only
final entries, but next-to-final entries; however, again these
constraints are satisfied automatically.

For both MHV and NMHV, the pentagon operator product expansion
(POPE)~\cite{Basso2013vsa,Basso2013aha,BelitskyI,POPE}
served previously as a powerful bootstrap
constraint~\cite{Dixon2013eka,Dixon2014iba}.
Now Regge exponentiation is enough to obtain a unique result.
Nonetheless, we do check our results against the POPE predictions.
We find complete agreement through five loops,
to each order in the OPE we have computed ($T^1$ and $T^2 F^2$ for
MHV~\cite{Basso2013vsa,Papathanasiou2013uoa,Basso2014koa,%
Papathanasiou2014yva,Drummond2015jea,YorgosPrivate}
and $T^1$ for the $(6134)$ component of
NMHV~\cite{Basso2013aha,Papathanasiou2013uoa}).

In ref.~\cite{Dixon2014iba}, two of the authors conjectured a relationship
between the $L$-loop MHV amplitude and the $(L{-}1)$-loop NMHV amplitude.
Our five-loop MHV amplitude allows us to verify this relation at one more
loop order. Expressed in terms of the functions
defined in eq.~(\ref{relations_to_standard}), it reads
(using the coproduct notation~\cite{Dixon2014iba})
\ba
 g^2\left(2E {-}  \EE\right)
 &=& \EE^{y_u,y_u}{+} \EE^{y_w,y_w} {-} 3 \EE^{y_v,y_v}{-} \EE^{v,v} {-} \EE^{1-v,v} 
   \nl
&&{+} 2 ( \EE^{y_u,y_v} {+} \EE^{y_w,y_v} ) {-} \EE^{y_u,y_w} {-} \EE^{y_w,y_u}.
\ea
This relation calls out for explanation.

Remarkably, the space of Steinmann functions appears to be 
``not much larger'' than required to contain $\EE$, $E$ and $\Et$,
if we include all derivatives of higher loop amplitudes.
Up to at least weight 6, the complete space is needed, apart from certain
unexpected restrictions on zeta values.  For example, the weight 2
functions found by taking 8 derivatives of $\EE^{(5)}$, $E^{(5)}$ and $\Et^{(5)}$
span a 6 dimensional subspace of the 7 dimensional Steinmann space:
$K_{1,1}^u$, $L_2^u + 2\zeta_2$, plus cyclic;
$\zeta_2$ is not an independent element.
In an ancillary file, we provide a coproduct representation
of this trimmed basis, which suffices to describe $\EE$, $E$ and $\Et$
through five loops. We also give HPL expressions for these functions
on the lines $(1,v,v)$ and $(u,1,1)$~\cite{SteinmannWebsite}.


\section{Conclusion}

Leveraging the power of the Steinmann relations, we have bootstrapped
six-point scattering amplitudes in planar $\mathcal{N}=4$ super Yang-Mills
theory through five loops.  Loop by loop, these amplitudes are dramatically
simpler than one would expect. Crucially, we did not need any external input:
all constraints imposed are either general or are fixed by behavior
at lower loops.  Yet higher loops, or even finite coupling, may well be
accessible too.

Unlike other techniques used to calculate in $\mathcal{N}=4$ SYM, the
Steinmann relations apply in general quantum field theories.
Their strength here suggests that these often-neglected constraints
may have broader applicability, perhaps making similar bootstrap techniques
viable in other theories, such as QCD.


{\it Acknowledgments.}
We thank Vittorio Del Duca, James Drummond, 
Claude Duhr, Georgios Papathanasiou
and Mark Spradlin for helpful discussions.
SCH's research was partly funded by the Danish National Research
Foundation (DNRF91). This research was supported by the US Department of Energy under contract DE--AC02--76SF00515, and by Perimeter Institute for Theoretical Physics. Research at Perimeter Institute is supported by the Government of Canada through the Department of Innovation, Science and Economic Development Canada and by the Province of Ontario through the Ministry of Research, Innovation and Science.  LD thanks the Kavli Institute for Theoretical Physics (National Science Foundation grant NSF PHY11-25915), and we thank Nordita, for hospitality.


\bibliographystyle{apsrev4-1}

\end{document}